\newcommand{\zboson}{Z}

\newcommand{\mz}{M_{Z}}

\newcommand{\mrec}{M_{\texttt{rec}}}
\newcommand{\mmu}{M_{\mu^{+}\mu^{-}}}
\newcommand{\ms}{M_{S}}
\newcommand{\fb}{\mathrm{fb}^{-1}}

\documentclass{PoS}

\title{Search for Extra Scalars Produced in Association with Muon Pairs at the ILC}

\ShortTitle{Search for Extra Scalars at the ILC}

\author{Yan Wang$^{\diamond *}$, Jenny List$^{\diamond}$, Mikael Berggren$^{\diamond}$\\
        $^{\diamond}$ DESY, Notkestra{\ss}e 85, 22607 Hamburg, Germany\\
        $^{*}$ IHEP, 19B Yuquan Road, Shijingshan District, Beijing, China\\
        E-mail: \email{yan.wang@desy.de}}

\author{~~~~~~on behalf of the International Large Detector concept group
}

\abstract{
We study the search for an extra scalar $S$ boson produced in association with the $\zboson$ boson at the International Linear Collider (ILC). The study is performed at center-of-mass energies of 250 GeV and 500 GeV based on the full simulation of the International Large Detector (ILD).  In order to be as model-independent as possible, the analysis uses the recoil technique, in particular with the $\zboson$ boson decaying into a pair of muons. As a result, exclusion cross-section limits are given in terms of a scale factor $k$ with respect to the Standard Model Higgs-strahlung process cross section. These predicted results, covering all possible searching regions of the extra scalars at the 250 GeV ILC and the 500 GeV ILC, can be interpreted independently of the decay modes of the $S$ boson.}

\FullConference{Talk presented at the International Workshop on Future Linear Colliders (LCWS2018),\\ Arlington, Texas, 22-26 October 2018.\\
 C18-10-22.}

\begin{document}
\section{Introduction}
The motivation of our study is to find a new scalar $S$ boson in the $SZZ$ coupling since one or more extra scalars are predicted in many new physics models. However, the properties of 125 GeV scalar measured
at the LHC is very similar to the Standard Model (SM) Higgs boson \cite{ATLAS}. As a result, the new scalar's coupling will be highly suppressed \cite{aggleton}. Furthermore, the LEP/LHC constraints on the extra
scalars always rely on the model details. Thus, a more precise analysis with model-independent assumptions to a scalar with the small coupling
is preferred. Although the OPAL collaboration has searched
for light scalars (less than 100 GeV) in a model-independent way at LEP, the results are limited by the low luminosity
\cite{Abbiendi}. The International Linear Collider (ILC) is a proposed electron-positron linear collider, whose luminosity will be over a thousand
times higher than that of LEP, which makes the recoil mass technique more accurate to find such extra scalars \cite{Asner}. And the ILC has higher center-of-mass energies, which will cover more searching regions for the extra scalar.
A preliminary version of this analysis has been reported at LCWS2017 \cite{Wang:2018fcw} and ICHEP2018 \cite{Wang:2018ichep}. Thus, only the updates from ICHEP2018 is summarized in this contribution.

\section{Event Generation and Detector Simulation}
The signal is $e^{+}e^{-}\to S+Z$ production, where the $\zboson$ boson
decays to a pair of muons. The decay branching ratios of $S$ are fixed as same as the 125 GeV Higgs boson, but no use would be
made of this fact.
As SM backgrounds, bremsstrahlung and initial state radiation (ISR) are explicitly considered for
all events. 
The event samples are generated with 100\% left-handed and right-handed beam polarization, using the Whizard 1.95 Monte Carlo (MC) event generator \cite{Kilian:2007gr}. Then the samples are reweighted with beam polarizations of $\pm 80\%$ for the electron beam and $\pm 30\%$ for the positron beam. 

The event samples are generated, simulated and reconstructed for different center-of-mass energies ($\sqrt{s}=$250 GeV and 500 GeV). 
In 250 GeV cases, we use the same setting as the samples generated in the context of the ILD Detailed Baseline Design document \cite{Abramowicz}. The fractions of integrated luminosity 2000 $\fb$ are dedicated
to the four sign combinations $(-+, +-, ++, --) = (45\%, 45\%, 5\%, 5\%)$. The signal benchmark points are chosen as every 5 GeV in the range of
$10 \leq \ms \leq 160$ GeV (totally 30 signal benchmark points). In 500 GeV cases, we use the samples generated in the context of the ILD Design Report \cite{IDR}, the fractions of integrated luminosity 4000 $\fb$ are dedicated
to $(-+, +-, ++, --) = (40\%, 40\%, 10\%, 10\%)$. Totally 48 signal benchmark points are chosen in the range of $10 \leq \ms \leq 410$ GeV.  Event reconstruction has been performed using the PandoraPFA algorithm to reconstruct individual final state particles, so-called Particle Flow Objects (PFOs).

\subsection{Event selection and Background Rejection}
Firstly, a pair of oppositely charged muons is selected by minimizing the following $\chi^{2}$ function:
\begin{equation}
\chi^{2} (\mmu, \mrec) =
\frac{(\mmu-\mz)^2}{\sigma^{2}_{\mmu}} + 
\frac{(\mrec-\ms)^2}{\sigma^{2}_{\mrec}},
\end{equation} 
where $\mmu$ and ${\mrec}$ are the invariant mass and the recoil mass of the muon pair, and $\sigma_{\mmu}$  and $\sigma_{\mrec}$ are determined by a Gaussian fit to the generator-level distributions of
$\mmu$ and ${\mrec}$. Then, the bremsstrahlung and final state radiation photons from the muon are combined with the muon.

\begin{figure}[ht]
    \begin{center}		
\includegraphics[height=8cm]{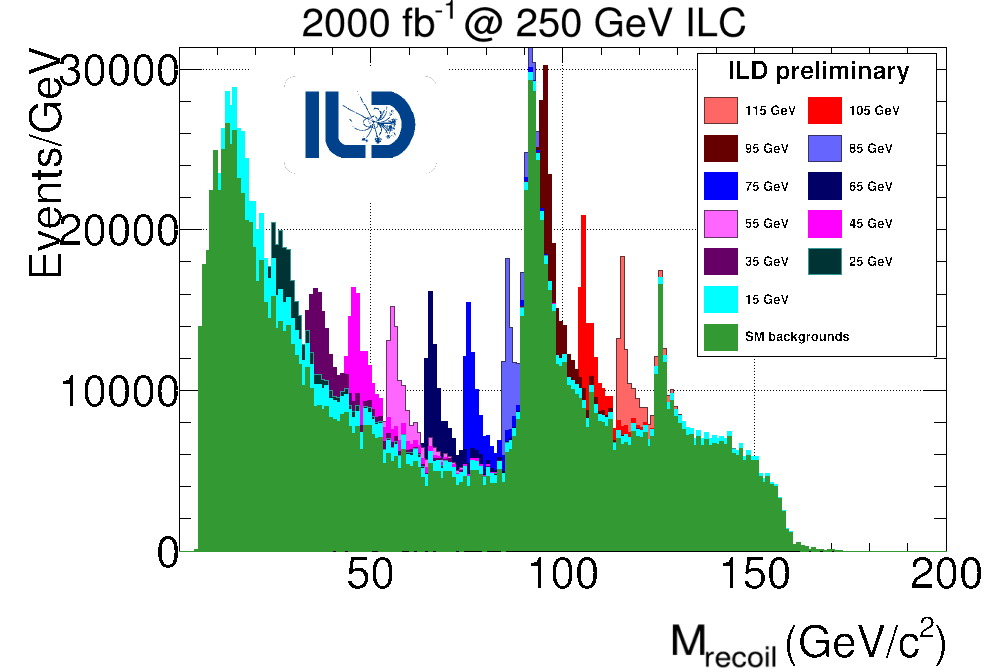} 
  \end{center}
  \vspace{-0.5cm}
  \caption{The recoil mass distributions for signal and backgrounds at the 250 GeV center of mass energy. These distributions are before the ISR veto cuts.}\label{recoil}
\end{figure}

Background events are rejected by firstly considering kinematic variables only relied on muons
(and the reconstructed $\zboson$ boson): the invariant mass and transverse momentum of the muon pair, as well
as the polar angle of the missing momentum. Then, a BDTG is trained using 6 input
variables based on TMVA \cite{MVA}: muon pair invariant mass, the polar angle of each muon, the polar angle of the muon pair, the opening angle of the muon pair, and the $\pi-(\phi_{\mu^{+}}-\phi_{\mu^{-}})$, where $\phi_{\mu^{\pm}}$ is the azimuthal angles of the muons with respect to the beam line. Finally, taking into account the ISR photon return effects, the two fermion background can be further rejected by ISR energy veto cuts.
With these cuts, no information on the decay of $S$ is needed, thus the expected results will be model-independent.
The recoil mass distributions are obtained after these cuts (Figure \ref{recoil}). 
\vspace{-0.5cm}

\section{Results}
A likelihood analysis is applied for calculating 2 $\sigma$ expected exclusion limits
on $k$ with a bin-by-bin comparison between the signal and background recoil mass histograms for each benchmark points, where $k$ is defined as 
\begin{equation}
k = \frac{\sigma_{SZ}}{\sigma_{H_{SM}Z}(m_{H_{SM}}=m_{S})},
\end{equation}
and $k_{95}$ is the 2 $\sigma$ exclusion limits for the cross section scale factor $k$ hereinafter.
\begin{figure}[]
    \begin{center}		
\includegraphics[height=8cm]{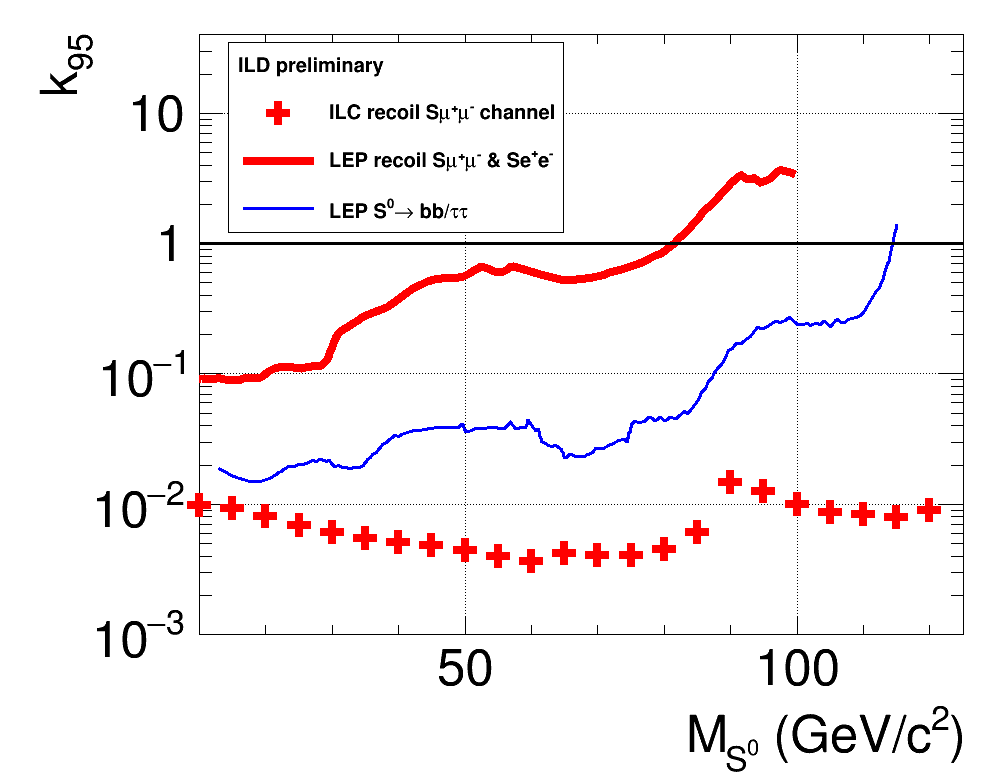} 
  \end{center}
  \caption{The directly comparison between the LEP and
ILC simulation results.  }\label{limits_exp}
\end{figure}

\begin{figure}[]
      \begin{center}
\includegraphics[height=8cm]{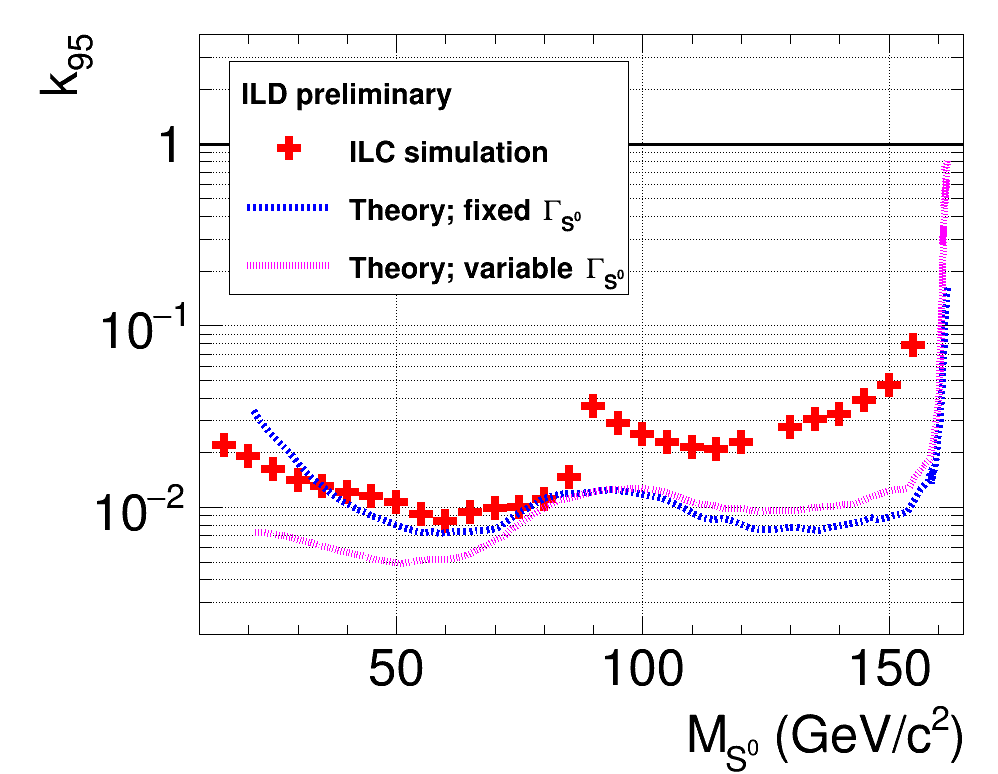} 
  \end{center}
  \caption{The comparison between the theoretical LEP extroplation results and
ILC expected results.}\label{limits_theo}
\end{figure}

In Figure \ref{limits_exp}, the ILC results at 250 GeV are compared with the LEP results directly. The red points are 2 $\sigma$ exclusion
limits for $\int Ldt = 2000~\fb$ and $\sqrt{S} = 250$ GeV at the ILC, while the red line was obtained with the
recoil mass method by the OPAL Collaboration \cite{Abbiendi} at LEP with about 0.8 $\fb$ data in total. Also shown with the blue line is the model-dependent results from LEP, combining measurements by ALEPH, DELPHI, L3, and OPAL \cite{Barate}, in which the decay modes of the scalars were utilized. In general, the ILC
exclusion limits will reach $10^{-2}$, and are one or two orders better than the OPAL recoil results and
even better than the LEP traditional results.

In Figure \ref{limits_theo},  the ILC theoretical predictions, which are extrapolated from the LEP measurements with fixed (variable) scalar width, are compared with the ILC simulation results \cite{theo}. The theoretical predictions combine $S\mu^{+}\mu^{-}$ and $Se^{+}e^{-}$ channels, while the ILC simulation results only use $S\mu^{+}\mu^{-}$ channel, but divide the results by $\sqrt{2}$~ ($k_{95}=k_{95}^{exp}(S\mu^{+}\mu^{-})/\sqrt{2}$).  These results are projected with 500 $\fb$ luminosity with $P(e^{-},e^{+})=(-80\%, +30\%)$.  From the figure, the ILC simulation results agree to the theoretical predictions with fixed scalar width in the low mass region.  However, the theoretical predictions extrapolate the expected background events in an interval around the $\zboson$ pole region \cite{theo}, so there is no $\zboson$ pole peak in the theoretical curves; at the same time, the theoretical predictions don't include SM Higgs background, as a result, they are better  than simulation results in the high mass region. 
\begin{figure}[ht]
    \begin{center}		
\includegraphics[height=8cm]{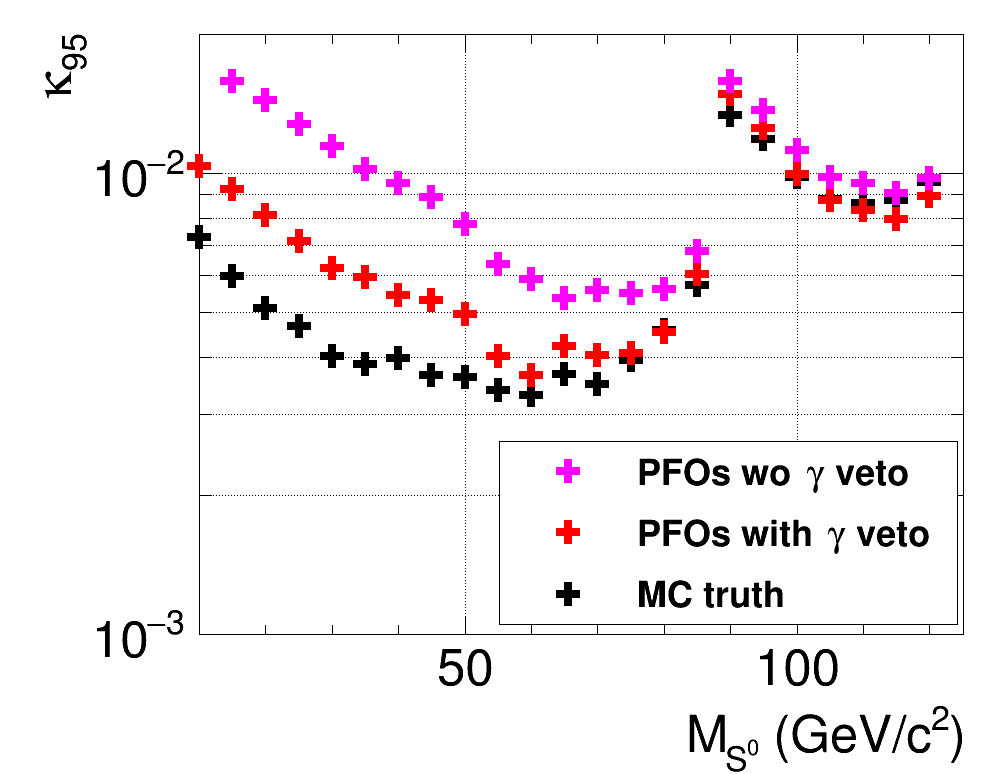} 
  \end{center}
  \caption{ The
effects for ISR photon veto cuts and the reconstruction efficiency.}\label{compare_de}
\end{figure}

\begin{figure}[ht]
    \begin{center}		
\includegraphics[height=8cm]{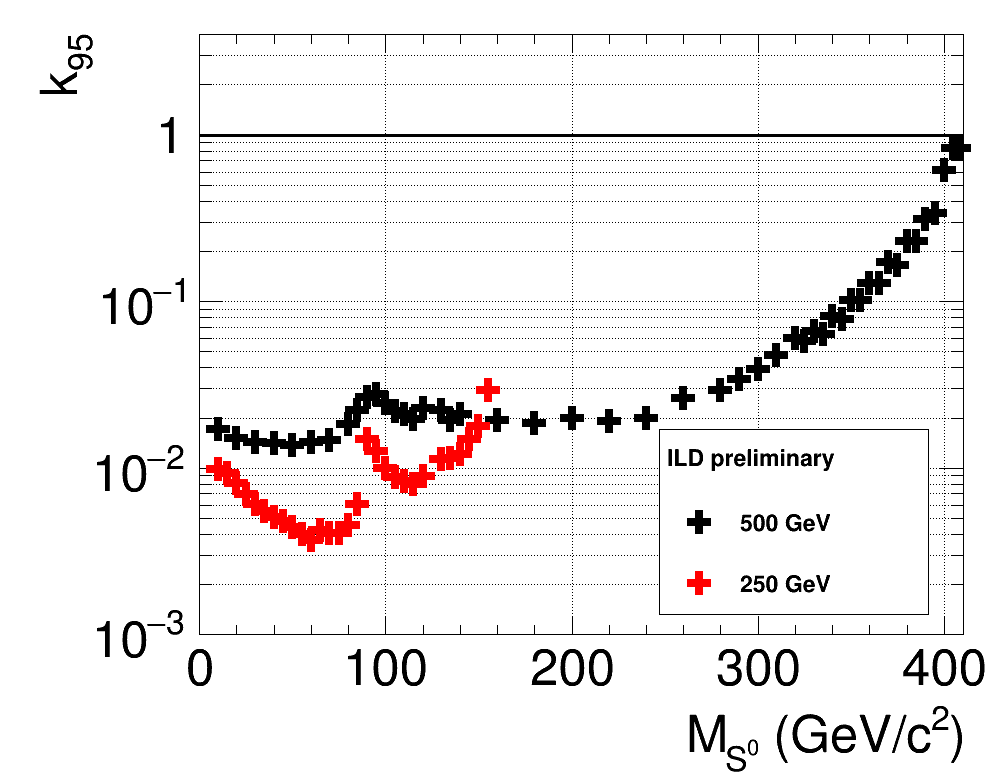} 
  \end{center}
  \caption{Preliminary Final Exclusion Limits for 250/500 GeV ILC.}\label{compare_500}
\end{figure}
In Figure \ref{compare_de}, a comparison is given among the PFO and their corresponding MC simulation inputs (MCtruth). The red and magenta points are the results with/without the ISR photon veto cuts
on the level of PFOs, because the two fermion background can be efficiently discarded by
considering the ISR effects. Thus, the ISR photon reconstructed efficiency will affect the result significantly. The black points are the results after photon veto cuts using MCtruths with the detector simulation, which reflect the best searching capability. And the difference between MCtruths and PFOs results shows we can improve the results with better photon reconstructions. 

In Figure. \ref{compare_500}, we show the preliminary exclusion limits for the 500 GeV ILC. In the low mass region, the 500 GeV results are worse than the 250 GeV cases mainly due to the suppressed cross sections, while they cover a larger searching region. Especially when $M_{S}<300$ GeV, $k_{95}$ is in the order of $10^{-2}$, which could set strong model-independent constraints for the extra scalars.

\section{Conclusions}
By applying the recoil technique, the potential of the ILC to search for scalars has been investigated at $\sqrt{S} = 250$ GeV and 500 GeV, with the full simulation of the ILD concept. The method is optimized to be independent of the scalar decay modes. $2~\sigma$ expected exclusion limits for the cross section scale factor $k_{95}$ are shown for scalar
mass from 10 GeV to 160 GeV when $\sqrt{S}=250$ GeV and from 10 GeV to 410 GeV when  $\sqrt{S}=500$. They are one or two orders of magnitudes more sensitive than LEP, and
covering substantial new phase spaces.

\section*{Acknowledgements}
We would like to thank the LCC generator working group and the ILD software
working group for providing the simulation and reconstruction tools and producing
the Monte Carlo samples used in this study. This work has benefited from
computing services provided by the ILC Virtual Organization, supported by the
national resource providers of the EGI Federation and the Open Science GRID. We
are grateful for the support from Collaborative Research Center SFB676 of the
Deutsche Forschungsgemeinschaft (DFG), Particles, Strings and the Early Universe,
project B1. Y.W. is supported by the China Postdoctoral Science Foundation
under Grant No. 2016M601134, and an International Postdoctoral Exchange
Fellowship Program between the Office of the National Administrative Committee
of Postdoctoral Researchers of China (ONACPR) and DESY.

\end{document}